\documentclass[aps,prb,twocolumn,showpacs,floatfix]{revtex4}

\usepackage{amsmath,amssymb,graphicx}

\begin{document}
\preprint{Submitted to Phys. Rev. B}


\title{Theory of carrier dynamics and time resolved reflectivity in
InMnAs/GaSb heterostructures}

\author{G. D. Sanders}
\author{C. J. Stanton}
\affiliation{Department of Physics, University of Florida, Box 118440\\
Gainesville, Florida 32611-8440}

\author{J. Wang}
\author{J. Kono}
\affiliation{Department of Electrical and Computer Engineering,
Rice Quantum Institute, and Center for Nanoscale Science and Technology,
Rice University, Houston, Texas 77005}

\author{A. Oiwa}
\author{H. Munekata}
\affiliation{Imaging Science and Engineering Laboratory,
Tokyo Institute of Technology, 4259 Nagatsuta, Yokohama 226-8503, Japan}

\date{\today}


\begin{abstract}
We present detailed theoretical calculations of two color,
time-resolved pump-probe differential reflectivity measurements. The
experiments modeled were performed on InMnAs/GaSb heterostructures
and showed pronounced oscillations in the differential reflectivity
as well as a time-dependent background signal. Previously, we showed
that the oscillations resulted from generation of coherent acoustic
phonon wavepackets in the epilayer and were not associated with the
ferromagnetism. Now we take into account not only the oscillations,
but also the background signal which arises from photoexcited
carrier effects. The two color pump-probe reflectivity experiments
are modeled using a Boltzmann equation formalism. We include
photogeneration of hot carriers in the InMnAs quantum well by a pump
laser and their subsequent cooling and relaxation by emission of
confined LO phonons. Recombination of electron-hole pairs via the
Schockley-Read carrier trapping mechanism is included in a simple
relaxation time approximation. The time resolved differential
reflectivity in the heterostructure is obtained by solving Maxwell's
equations and compared with experiment. Phase space filling, carrier
capture and trapping, band-gap renormalization and induced
absorption are all shown to influence the spectra.

\end{abstract}



\maketitle

\section{Introduction}
\label{Introduction}

Dilute magnetic semiconductors made of (III,Mn)V materials have
recently garnered much attention owing to the discovery of carrier
mediated ferromagnetism.  This offers the intriguing  possibility of
their use in the development of semiconductor spintronic devices
capable of simultaneously performing information processing, data
storage, and communication functions. \cite{Zutic04.323, Ohno98.951,
Ohno01.840}

Time-independent optical studies such as cyclotron resonance of
III-V DMS materials have provided important information on effective
masses and electronic structure,\cite{Zudov02.161307} exchange
parameters,\cite{Sanders03.6897, Sanders03.165205} carrier
densities,\cite{Sanders04.378} and whether carriers are localized or
itinerant.\cite{Matsuda04.195211}

Time-dependent optical studies are even more useful and can provide
information that static magnetization or electrical transport
measurements can not.  In this paper, we report on our theoretical
calculations and modeling of femtosecond,  time-resolved
differential reflectivity spectroscopy of InMnAs/GaSb
heterostructures. Femtosecond transient reflectivity spectroscopy
has proven useful in studying carrier dynamics in semiconductors as
well as the generation and propagation of coherent phonons in a
number of materials. In particular, coherent optical phonons have
been observed in bulk semiconductors \cite{Dekorsey93.3842,
Kuznetsov95.7555} and coherent acoustic phonons have been detected
in InGaN/GaN-based semiconductor heterostructures.
\cite{Chern04.339,Yahng02.4723,Stanton03.525,Liu03.0310654}

The experiments we model are two-color pump-probe differential
reflectivity measurements.  In these experiments, there are several
changes to the reflectivity on different time-scales. On the fast
time scale, there are changes to the reflectivity associated with
ultrashort carrier lifetimes ($\sim $ 2 ps) and multi-level
dynamics.

In addition, pronounced oscillations are observed on a longer time
scale ($\sim$23 ps period).\cite{Wang05.0505261, Wang05.accepted}
Similar behavior was seen in InGaMnAs systems.\cite{Wang03.563}
Originally these oscillations were thought to be associated with the
ferromagnetism in the InMnAs layer since oscillations were not
observed in samples without Mn.  However, we showed in a previous
work\cite{Wang05.accepted} that the oscillations instead resulted
from selective photoexcitation in the InMnAs layer which triggered a
coherent phonon wavepacket that propagated into the GaSb layer.

In this paper, we expand upon our previous work and show that the
large strength of the coherent phonon oscillations results from a
fortuitous strong dependence on strain of the GaSb dielectric
function near the probe energy. In addition, we also model the fast
time dependent background signal which arises from the photoexcited
carriers. There are three main contributions to this
transient background signal: 1) the enhanced Drude absorption
resulting from the photoinduced increase in free carriers,
2) the relaxation dynamics associated with the decay of the highly
nonequilibrium photoexcited carrier distribution and 3) the trapping
and subsequent non-radiative recombination of photoexcited carriers
due to the high density of defects in the InMnAs layer.

We model the experiments by first calculating the detailed
electronic structure in the InMnAs layer.  We than use a Boltzmann
equation formalism to account for the photoexcited carrier dynamics.
We include band-gap renormalization, carrier-phonon scattering, and
carrier-trapping/recombination through the Shockley-Read mechanism.
To determine the optical properties, we solve Maxwell's equations in
the heterostructure. Details of the experiments and the calculations
are given in the following sections.

\section{Experiment}
\label{Experiment}

\begin{figure}[tbp]
\includegraphics[scale=0.25]{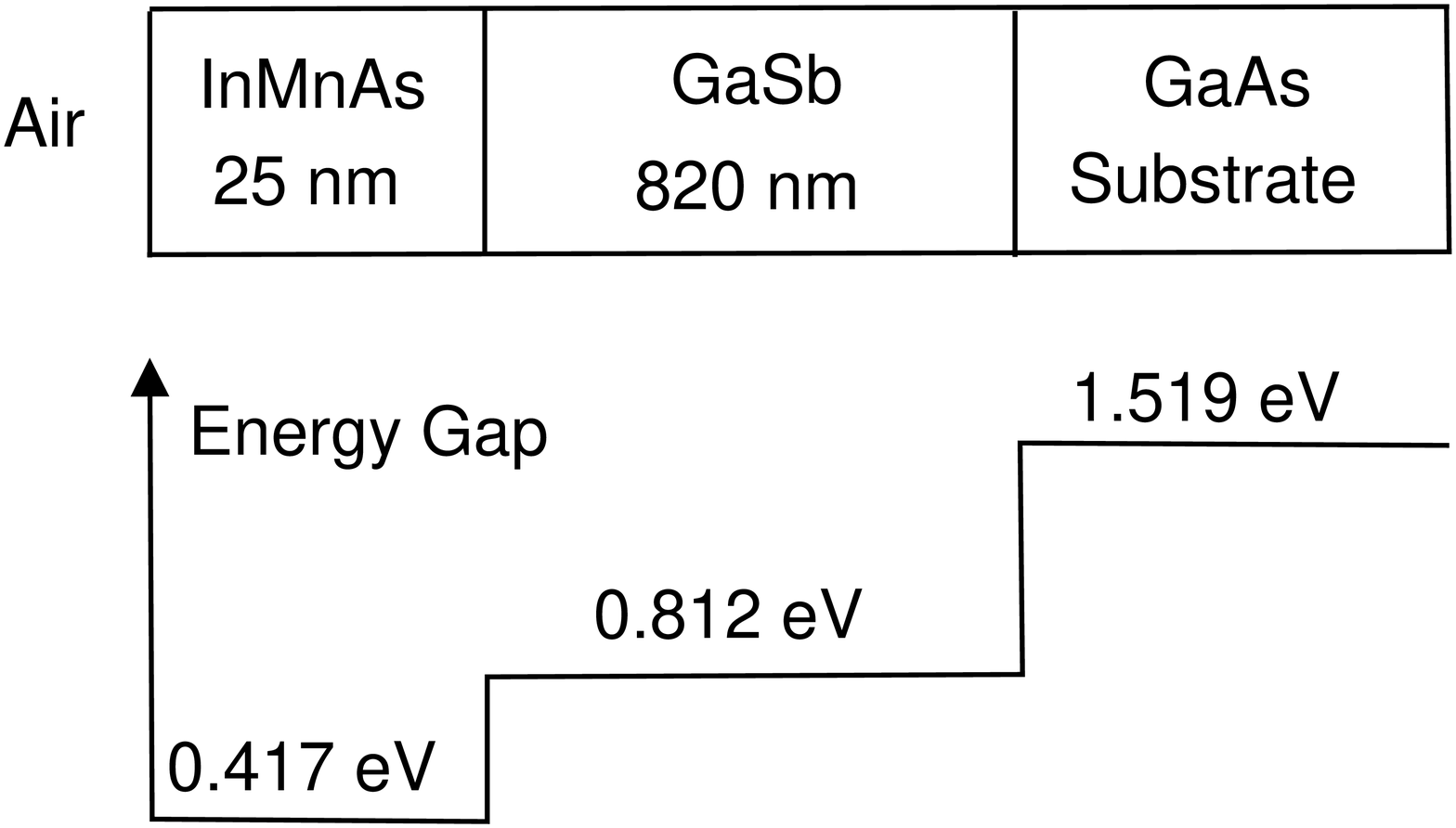}
\caption{Schematic diagram of the
heterostructure consisting of a 25 nm In$_{0.91}$Mn$_{0.09}$As quantum
well and a 820 nm GaSb barrier grown on a GaAs substrate. The band
gap as a function of position in each of the layers is also
shown.}
\label{Structure}
\end{figure}

The main sample studied (shown schematically in Fig.~\ref{Structure})
was an InMnAs/GaSb heterostructure, consisting of a 25 nm thick
magnetic layer with Mn concentration 0.09, grown on a 820 nm thick GaSb
buffer layer on a semi-insulating GaAs (001) substrate.  Its room temperature
hole density and mobility were 1.1 $\times$ 10$^{19}$ cm$^{-3}$ and
323 cm$^2$/Vs, respectively, estimated from Hall measurements.
Detailed growth conditions and sample information can be
found in Ref.~\onlinecite{Slupinski02.1326}.

We performed two-color time-resolved differential reflectivity spectroscopy
using femtosecond midinfrared pump pulses (2 $\mu$m, $\sim$ 140 fs) and a white
light continuum probe (0.5 - 1.4 $\mu$m, $\sim$ 340 fs).
Experimental details are described in Ref.~\onlinecite{Wang04.2771}.

The source of intense MIR pulses was an optical parametric amplifier (OPA) pumped by
a Ti:Sapphire-based regenerative amplifier
(Model CPA-2010, Clark-MXR, Inc., 7300 West Huron River Drive, Dexter, MI 48130).
At the pump wavelength (2 $\mu$m), the photon energy (0.62 eV)
was just above the band gap of InMnAs, so the created carriers had only a
small amount of extra kinetic energy ($\sim$ 0.2 eV at 15 K), minimizing
contributions from intervalley scattering and intraband relaxation. A white light
continuum generated by focusing a small fraction of the CPA pulses into a
sapphire crystal was used as a probe, which allowed us to probe a wide energy
range far above the quasi-Fermi level of the optically excited carriers.

\section{Theory}
\label{Theory}

\subsection{Bulk InMnAs Bandstructure}

In our pump-probe reflectivity experiments
carriers are created in the InMnAs layer by pumping below
the GaSb bandgap. We treat the photogenerated carriers (electrons
and holes) in an 8 band $k \cdot p$ effective mass model which
includes conduction electrons, heavy holes, light holes,
and split-off holes. Following Pidgeon and Brown \cite{Pidgeon66.575}
we find it convenient to separate the 8 Bloch basis states into
upper and lower sets which decouple at the zone center. The Bloch basis
states for the upper set are
\begin{subequations}
\label{upperset}
\begin{eqnarray}
\arrowvert 1 \rangle &=&
\arrowvert \frac{1}{2}, +\frac{1}{2} \ \rangle=
\arrowvert S \uparrow \rangle
\\
\arrowvert 2 \rangle &=&
\arrowvert \frac{3}{2}, +\frac{3}{2} \ \rangle=
\frac{1}{\sqrt{2}} \arrowvert (X + i Y) \uparrow \rangle
\\
\arrowvert 3 \rangle &=&
\arrowvert \frac{3}{2}, -\frac{1}{2} \rangle=
\frac{1}{\sqrt{6}} \arrowvert (X - i Y) \uparrow +2 Z \downarrow \rangle
\\
\arrowvert 4 \rangle &=&
\arrowvert \frac{1}{2}, -\frac{1}{2} \rangle=
\frac{i}{\sqrt{3}} \arrowvert -(X - i Y) \uparrow +Z \downarrow \rangle
\end{eqnarray}
\end{subequations}
which correspond to electron spin up, heavy hole spin up,
light hole spin down, and split off hole spin down.
Likewise, the Bloch basis states for the lower set are
\begin{subequations}
\label{lowerset}
\begin{eqnarray}
\arrowvert 5 \rangle &=&
\arrowvert \frac{1}{2},-\frac{1}{2} \rangle=
\arrowvert S \downarrow \rangle
\\
\arrowvert 6 \rangle &=&
\arrowvert \frac{3}{2}, -\frac{3}{2} \ \rangle=
\frac{i}{\sqrt{2}} \arrowvert (X - i Y) \downarrow \rangle
\\
\arrowvert 7 \rangle &=&
\arrowvert \frac{3}{2}, +\frac{1}{2} \rangle=
\frac{i}{\sqrt{6}} \arrowvert (X + i Y) \downarrow -2 Z \uparrow \rangle
\\
\arrowvert 8 \rangle &=&
\arrowvert \frac{1}{2}, +\frac{1}{2} \rangle=
\frac{1}{\sqrt{3}} \arrowvert (X + i Y) \downarrow +Z \uparrow \rangle
\end{eqnarray}
\end{subequations}
corresponding to electron spin down, heavy hole spin down, light
hole spin up, and split off hole spin up.

The effective mass Hamiltonian in bulk zinc blende materials
in the axial approximation is given explicitly by \cite{Efros88.7120}
\begin{equation}
H_0= \left[
\begin{array}{cc}
H_{uu} & H_{ul} \\
H_{lu} & H_{ll} \\
\end{array}
\right]
\label{H0}
\end{equation}
where $H_{uu}$, $H_{ul}$, $H_{lu}$, and $H_{ll}$ are $4 \times 4$
submatrices. The effective mass Hamiltonian matrix elements between
the upper set basis states in Eq.~(\ref{upperset}) are
\begin{subequations}
\label{H0 submatrices}
\begin{equation}
H_{uu}=\left[
\begin{array}{cccc}
E_g+A & i \frac{\sqrt{2}}{2} V k_{+} & i \frac{\sqrt{6}}{6} V k_{-} &
\frac{\sqrt{3}}{3} V k_{-} \\
-i \frac{\sqrt{2}}{2} V k_{-} & -P-Q & -M & i \sqrt{2} M \\
-i \frac{\sqrt{6}}{6} V k_{+} & -M^\dag & -P+Q & i \sqrt{2} Q \\
\frac{\sqrt{3}}{3} V k_{+} & -i \sqrt{2} M^\dag & -i \sqrt{2} Q & -\Delta-P \\
\end{array}
\right],
\end{equation}
while the Hamiltonian matrix elements between the lower set basis states
in Eq.~(\ref{lowerset}) are given by
\begin{equation}
H_{ll}=\left[
\begin{array}{cccc}
E_g+A & -\frac{\sqrt{2}}{2} V k_{-} & -\frac{\sqrt{6}}{6} V k_{+} &
i \frac{\sqrt{3}}{3} V k_{+} \\
-\frac{\sqrt{2}}{2} V k_{+} & -P-Q & -M^\dag & i \sqrt{2} M^\dag \\
-\frac{\sqrt{6}}{6} V k_{-} & -M & -P+Q & i \sqrt{2} Q \\
-i\frac{\sqrt{3}}{3} V k_{-} & -i\sqrt{2} M & -i\sqrt{2} Q & -\Delta-P \\
\end{array}
\right].
\end{equation}
The submatrices coupling upper and lower set basis states are
\begin{equation}
H_{ul}=\left[
\begin{array}{cccc}
0 & 0 & \frac{\sqrt{6}}{3} V k_z & i \frac{\sqrt{3}}{3} V k_z \\
0 & 0 & -L & -i \frac{\sqrt{2}}{2} L \\
-i \frac{\sqrt{6}}{3} V k_z & L & 0 & i \frac{\sqrt{6}}{2} L^\dag \\
-\frac{\sqrt{3}}{3} V k_z & -i \frac{\sqrt{2}}{2} L & i \frac{\sqrt{6}}{2} L^\dag & 0 \\
\end{array}
\right]
\end{equation}
and
\begin{equation}
H_{lu}=\left[
\begin{array}{cccc}
0 & 0 & i \frac{\sqrt{6}}{3} V k_z & -\frac{\sqrt{3}}{3} V k_z \\
0 & 0 & L^\dag & i\frac{\sqrt{2}}{2} L^\dag \\
\frac{\sqrt{6}}{3} V k_z & -L^\dag & 0 & -i \frac{\sqrt{6}}{2} L \\
-i\frac{\sqrt{3}}{3} V k_z & i\frac{\sqrt{2}}{2} L^\dag &
-i\frac{\sqrt{6}}{2} L & 0 \\
\end{array}
\right].
\end{equation}
\end{subequations}
In Eqs.~(\ref{H0 submatrices}) $E_g$ is the bulk band gap,
$\Delta$ is the spin orbit splitting and $k_{\pm} = k_x \pm i \ k_y$.
Following Jain \textit{et al} \cite{Jain90.3747},
band gap renormalization in InAs shrinks the
InAs band gap by an amount
\begin{equation}
\Delta E_g = A \ N^{1/3} + B \ N^{1/4} + C \ \sqrt{N}
\label{band gap renormalization}
\end{equation}
where $N$ is the electron or hole carrier concentration and
$A$, $B$, and $C$ are  material parameters which are found to be
different for electrons and holes.
The temperature dependence of the band gap is taken into
account using the empirical Varshni formula \cite{Varshni67.149}.
The temperature and carrier dependent band gap is given by
\begin{equation}
E_g(T,N) = E_g - \frac{\alpha_v \ T^2}{T + \beta_v}-\Delta E_g
\label{Varshni formula}
\end{equation}
where $\alpha_v$ and $\beta_v$ are Varshni parameters which
are tabulated in Ref. \onlinecite{Vurgaftman01.5815} for
a variety of semiconductors.  The Kane momentum matrix element,
$V = (-i \hbar/m_0) \langle S \arrowvert p_x \arrowvert X \rangle $,
is related to the optical matrix element $E_p$
by \cite{Vurgaftman01.5815}
\begin{equation}
V = \sqrt{ \frac{\hbar^2}{m_0} \frac{E_p}{2} }.
\label{Kane momentum matrix element}
\end{equation}
The operators $A$, $P$, $Q$, $L$ and $M$ are
\begin{subequations}
\label{APQLM}
\begin{equation}
A =\frac{\hbar^2}{m_0} \frac{\gamma_4}{2} \left(
k_\parallel^2+k_z^2 \right),
\end{equation}
\begin{equation}
P = \frac{\hbar^2}{m_0} \frac{\gamma_1}{2}
\left( k_\parallel^2 + k_z^2 \right),
\end{equation}
\begin{equation}
Q = \frac{\hbar^2}{m_0} \ \frac{\overline{\gamma}}{2}
\left( k_\parallel^2 - 2 k_z^2 \right),
\end{equation}
\begin{equation}
L = -i \frac{\hbar^2}{m_0} \sqrt{3} \ \overline{\gamma}
\left(k_x - i k_y \right) k_z,
\end{equation}
and
\begin{equation}
M = \frac{\hbar^2}{m_0} \frac{\sqrt{3}}{2} \ \overline{\gamma}
\left( k_x-i k_y \right)^2
\end{equation}
\end{subequations}
where $k_\parallel^2=k_x^2+k_y^2$. In the axial approximation
\cite{Baldereschi73.2697}, we have
$\overline{\gamma} = (2 \gamma_2+3 \gamma_3)/5$
so that the energy bands depend only on the magnitude of
$\mathbf{k}_\parallel$.
Note that at $k_z=0$, the effective mass Hamiltonain
in Eq.~(\ref{H0}) is block diagonal since $L=0$ at $k_z=0$.

In Eq. (\ref{APQLM}), the parameters $\gamma_1$ and $\overline{\gamma}$
are not the usual Luttinger parameters since this is an eight-band\
model, but instead are related to the usual Luttinger parameters
$\gamma_1^L$ and $\overline{\gamma}^L$
by the relations \cite{Luttinger56.1030}
\begin{equation}
\gamma_1 = \gamma_1^L - \frac{E_p}{3 E_g}
\end{equation}
and
\begin{equation}
\overline{\gamma} = \overline{\gamma}^L - \frac{E_p}{6 E_g}
\end{equation}
This takes into account the additional coupling of the
valence bands to the conduction band not present in the
six band Luttinger model for the valence bands.

The parameter $\gamma_4$ is related to the conduction-band
effective mass $m_e^*$ through the relation \cite{Efros88.7120}
\begin{equation}
\gamma_4 = \frac{1}{m_e^*} - \frac{E_p}{3}
\left(
\frac{2}{E_g} + \frac{1}{E_g + \Delta}
\right)
\end{equation}

The exchange interaction between the Mn$^{++}$ $d$ electrons and
the conduction $s$ and valence $p$ electrons is treated in the
virtual crystal and molecular field approximation. The resulting
Mn exchange Hamiltonian is \cite{Kossut88.183}
\begin{equation}
H_{Mn}= x\ N_0\ \langle S_z \rangle \left[
\begin{array}{cc}
D_a & 0 \\
0   & -D_a
\end{array}
\right] \label{H_Mn}
\end{equation}
where $x$ is the Mn concentration, $N_0$ is the number of cation
sites in the sample, and $\langle S_z \rangle$ is the average spin
on a Mn site. The $4 \times 4$ submatrix $D_a$ is
\begin{equation}
D_a= \left[
\begin{array}{cccc}
\frac{1}{2}\alpha & 0 & 0 & 0 \\
0 & \frac{1}{2}\beta & 0 & 0 \\
0 & 0 & -\frac{1}{6}\beta & -i\frac{\sqrt{2}}{3}\beta \\
0 & 0 & i\frac{\sqrt{2}}{3}\beta & \frac{1}{2}\beta
\end{array}
\right] \label{D_a}
\end{equation}
where $\alpha$ and $\beta$ are the
$s$-$d$ and $p$-$d$ exchange integrals.

The average Mn spin $\langle S_z \rangle$ in the ferromagnetic
In$_{1-x}$Mn$_x$As quantum well is computed in the mean field
approximation, i.e.
\begin{equation}
\langle S_z \rangle = -S\ B_S
\left( - \frac{J S \langle S_z \rangle}{kT}
\right),
\end{equation}
where $B_S(x)$ is the Brillouin function, $S=5/2$ for the $3d^5$
electrons of the the Mn$^{++}$ ion \cite{Furdyna88.29},
$J=3 k_B T_c/S(S+1)$ is the ferromagnetic coupling, and $T_c$
is the experimentally measured Curie temperature.

The effective mass Hamiltonian for In$_{1-x}$Mn$_{x}$As is
\begin{equation}
H = H_0 + H_{Mn}.
\label{H_Total}
\end{equation}
It is assumed in our calculations that the compensation arises
from As antisites and hence the effective Mn fraction, $x$, in Eq.
(\ref{H_Mn}) is taken to be equal to the actual Mn fraction in the
sample. We note that this is supported by experimantal evidence
showing that InAs grown at low temperature (200 $^{\circ}$C) is a
homogeneous alloy and that the magnetization varies linearly with
Mn content, $x$.
\cite{Munekata89.1849,Munekata90.176,Ohno91.6103,Molnar91.356}

\subsection{Confined states in the InMnAs quantum well}

In quantum-confined systems such as the InMnAs quantum well shown in
Fig. \ref{Structure}, we must modify the bulk Hamiltonian in
Eq. (\ref{H_Total}). The quantum well breaks translational symmetry
along the $z$-direction but not in the $x$-$y$ plane.
Since the pump pulse is below the GaSb band gap, all
photogenerated electrons and holes are strongly confined to the well
and we assume the confinement potentials are infinite.
The wave functions in the envelope function approximation are
\begin{equation}
\psi_{n,\mathbf{k}_\parallel}(r) =
\frac{e^{i \mathbf{k}_\parallel \cdot \mathbf{\rho}} }{ \sqrt{A} }
\sum_{\nu=1}^8 F_{n,\nu,\mathbf{k}_\parallel} (z)
\ \arrowvert \nu \rangle
\end{equation}
where $A$ is the cross sectional area of the sample, $n$ is the subband
index, $\mathbf{k}_\parallel=(k \cos \theta, \ k \sin \theta, \ 0)$
is the two dimensional wave vector and
$\arrowvert \nu \rangle=\arrowvert 1 \rangle \ldots \arrowvert 8 \rangle$
are the Bloch basis states defined in
Eqs. (\ref{upperset}) and (\ref{lowerset}). The complex valued envelope
functions $F_{n,\nu,\mathbf{k}_\parallel}(z)$ are slowly varying in
comparison with the Bloch basis states.

The envelope functions satisfy a set of effective-mass
Schr\"{o}dinger equations which, in the axial approximation, are
\begin{equation}
\sum_{\nu'=1}^8 H_{\nu,\nu'}(\mathbf{k}_\parallel)
\ F_{n,\nu',\mathbf{k}_\parallel} (z) =
E_n(k) \ F_{n,\nu,\mathbf{k}_\parallel} (z)
\label{Schrodinger}
\end{equation}
subject to the boundary condition that the envelope functions
vanish at the walls of the quantum well. The operators
$H_{\nu,\nu'}(\mathbf{k}_\parallel)$ depend on the wave vector in the
x-y plane and can be obtained from the matrix elements in
Eq. (\ref{H_Total}) by making the operator replacement
$k_z \rightarrow -i \ \partial /\partial z$ in all the matrix
elements of Eq. (\ref{H0}).

In practice, we solve for the envelope functions
for a given value of $\mathbf{k}_\parallel$ on an evenly spaced
mesh of points, $z_i, \ i=1 \ldots N$, in the quantum well.
Approximating the derivative, $\partial/\partial z$ by a finite
difference formula, the  Schr\"{o}dinger equation (\ref{Schrodinger})
with the rigid wall boundary conditions becomes a matrix eigenvalue
problem which can be solved for the eigenvalues
$E_n(\mathbf{k}_\parallel)$ and the complex envelope functions
$F_{n,\nu,\mathbf{k}_\parallel} (z_i)$ evaluated at the mesh
points, $z_i$.

\subsection{Boltzmann transport equations}

In the two-color time resolved differential reflectivity experiments
the pump laser is used to promote electrons from the valence to the
conduction subbands of the quantum well. The photoexcited
carriers then relax through scattering, changing the optical properties
of the heterostructure in the process. These processes are often
simulated using Boltzmann transport equations.

In this paper, we formulate and solve the Boltzmann transport equations
using a numerical method similar to the one described in
Ref. \onlinecite{Sanders98.9148}.
For each subband state with energy, $E_n(k)$, we have
a time dependent distribution function, $f_n(\mathbf{k}_\parallel,t)$,
which gives the probability, as a function of time, of finding an
electron in subband $n$ with wave vector $\mathbf{k}_\parallel$.
The Boltzmann equation including photogeneration of hot electron-hole
pairs by the pump, the subsequent cooling of these carriers by emission
of confined LO phonons, and the recombination of electron-hole pairs by
means of carrier trapping is
\begin{eqnarray}
\nonumber && \frac{\partial f_n(\mathbf{k}_\parallel)}{\partial t}
= \sum_{n',k_\parallel'} \{ f_{n'}(\mathbf{k}_\parallel')
\ W_{\mathbf{k}_\parallel',\mathbf{k}_\parallel}^{n',n}
\ \left[ 1-f_n(\mathbf{k}_\parallel) \right] \\
&& - f_n(\mathbf{k}_\parallel)
\ W_{\mathbf{k}_\parallel,\mathbf{k}_\parallel'}^{n,n'}
\ \left[ 1-f_{n'}(\mathbf{k}_\parallel') \right] \}
+ \left[ \frac{\partial f_n(\mathbf{k}_\parallel)}{\partial t} \right]
\label{2D Boltzmann equation}
\end{eqnarray}
The scattering rate due to scattering by confined LO phonons in
the quantum well,
$W_{\mathbf{k}_\parallel,\mathbf{k}_\parallel'}^{n,n'}$,
is the rate at which electrons in subband $n$ with wave vector
$\mathbf{k}_\parallel$ scatter to subband $n'$ with wave vector
$\mathbf{k}_\parallel'$. The last term on the right hand side of
Eq. (\ref{2D Boltzmann equation}) describes the change
in the electron distribution function due to the action of the pump
as well as recombination of electron-hole pairs by means of carrier
trapping. Thus
\begin{equation}
\left[ \frac{\partial f_n(\mathbf{k}_\parallel)}{\partial t} \right]=
\left[ \frac{\partial f_n(\mathbf{k}_\parallel)}{\partial t} \right]_P+
\left[ \frac{\partial f_n(\mathbf{k}_\parallel)}{\partial t} \right]_T.
\label{Photogeneration and Relaxation Terms}
\end{equation}
where the first term on the right hand side is the photogeneration
rate and the second term is the recombination rate due to carrier
trapping.

To simplify the calculations, we use an axial approximation in which
the distribution functions are replaced by their angular averages in
the $x$-$y$ plane of the quantum well. The axial distribution functions
are
\begin{equation}
f_n(k,t)=
\int_{- \pi}^{\pi} \frac{d \theta}{2\pi} \ f_n(\mathbf{k}_\parallel,t)=
\int_{- \pi}^{\pi} \frac{d \theta}{2\pi} \ f_n(k,\theta,t).
\end{equation}
Next, we divide $k$ space into evenly spaced
cells of width $\Delta k = k_{max}/N_k$ where $N_k$ is the number
of cells. The value of k at the midpoint of each cell is denoted
$k_m$ ($m=1 \ldots N_k$). In each $k$ cell we define the
cell averaged distribution functions
\begin{equation}
f_n(k_m,t)=\int_m \frac{dk \ k}{\Delta k \ k_m} \ f_n(k,t)
\end{equation}
where the limits of integration are from $k_m - \Delta k/2$
to $k_m + \Delta k/2$.

If we assume the distribution functions in the Boltzmann equation
depend only on $k$ and vary slowly within each $k$ cell, we can obtain
a coupled set of ordinary differential equations for the cell
averaged axial distribution functions
\begin{widetext}
\begin{equation}
\frac{\partial f_n(k_m)}{\partial t}
=\frac{A}{2 \pi} \sum_{n',m'} \Delta k \ k_m'
\{ f_{n'}(k_m') \ W_{m',m}^{n',n}
\ \left[ 1-f_n(k_m) \right] \\
- f_n(k_m) \ W_{m,m'}^{n,n'}
\ \left[ 1-f_{n'}(k_m') \right] \}
+ \left[ \frac{\partial f_n(k_m)}{\partial t} \right]
\label{1D Boltzmann equation}
\end{equation}
\end{widetext}
The confined LO phonon scattering rates in
the original Boltzmann equation (\ref{2D Boltzmann equation}) depend only
on the angle $\Theta$ between $\mathbf{k}_\parallel$ and
$\mathbf{k}_\parallel'$, so the cell and axially averaged scattering
rates appearing in Eq. (\ref{1D Boltzmann equation}) are given by
\begin{equation}
W_{m,m'}^{n,n'} =
\int_{m} \frac{dk \ k}{\Delta k \ k_m}
\int_{m'} \frac{dk' \ k'}{\Delta k \ k_{m'}}
\int_{-\pi}^{\pi} \frac{d \Theta}{2\pi} \
W^{n,n'}_{\mathbf{k}_\parallel,\mathbf{k}_\parallel'}
\label{Cell averaged scattering rates}
\end{equation}

Given initial values of the distribution function,
$f_n(k_m,t \rightarrow -\infty)$, we can solve the system of
coupled ordinary differential equations (\ref{1D Boltzmann equation})
with an adaptive Runge-Kutta routine. \cite{Press}
In order for the integration to be numerically stable,
we need the cell averaged scattering rates in
Eq. (\ref{Cell averaged scattering rates}) to satisfy the detailed
balance condition. We insure this by calculating the downward
scattering rate and using the detailed balance
condition to obtain the upward scattering rate. If
$W_{m,m'}^{n,n'}$ is the cell averaged downward scattering
rate between a higher lying state $E_n(k_m)$ and a lower lying state
$E_{n'}(k_{m'})$, then the upward scattering rate,
$W_{m',m}^{n'm}$, between these two states is
\begin{equation}
W_{m',m}^{n',n}=\exp \left[
- \left( \frac{E_n(k_m)-E_{n'}(k_{m'})}{k_B T}
\right) \right] \ W_{m,m'}^{n,n'}.
\label{Detailed balance}
\end{equation}

The electrons are initially in thermal equilibrium with the
lattice and are described by a Fermi-Dirac distribution
\begin{equation}
f_n^0(k_m)=\frac{1}{1+\exp\{ [E_n(k_m)-E_F]/k_B T \}}.
\label{Initial distribution function}
\end{equation}

If $n$ and $p$ are the initial electron and hole column densities
in the quantum well, then the Fermi energy as a function
of the lattice temperature and initial carrier concentrations
can be easily found by solving
\begin{equation}
n-p= \frac{1}{2\pi} \sum_{n,m} \Delta k \ k_m
\left( f_n^0(k_m) - \delta_{n,v} \right)
\label{Fermi Energy equation}
\end{equation}
for $E_F$ using a root finding routine. \cite{Press}
In Eq. (\ref{Fermi Energy equation})
the valence delta function $\delta_{n,v}$ is defined
to be zero if subband $n$ is a conduction subband and one if subband
$n$ is a valence subband.

To finish specifying the transport problem, we need to supply the
the cell averaged axial scattering rates appearing in
Eq. (\ref{1D Boltzmann equation}) as well as the cell averaged
photogeneration and carrier trapping term.

\subsection{Photogeneration rates}

The photogeneration rate is computed using Fermi's golden rule. The pump
is characterized by the fluence which is the total flux integrated over
time. Assuming a narrow spectral width for the pump, centered on the pump
energy $\hbar\omega$, the fluence is given by
\begin{equation}
F_0 = \int_{-\infty}^{\infty} U_0(t) \frac{c}{n_r}
\end{equation}
where $U_0(t)$ is the pump energy density and $n_r$ is the refractive
index in the InMnAs quantum well. For the pump energy density, we
assume a Gaussian pulse shape with an intensity FWHM of $\tau_p$ so
that
\begin{equation}
U_0(t)=\bar{U}_0 \exp \left(
-4 \ \ln(2) \left( \frac{t}{\tau_p} \right)^2 \right)
\end{equation}

Assuming the Dirac delta function in Fermi's golden rule is the only
rapidly varying quantity in a k-cell, the cell averaged photogeneration
rate in Eq. (\ref{Photogeneration and Relaxation Terms}) is given by
\begin{eqnarray}
\nonumber &&
\left[ \frac{\partial f_n(k_m)}{\partial t} \right]_P =
\frac{4\pi^2 \ e^2 \ U_0(t)}{\hbar \ n_r^2 \ (\hbar\omega)^2}
\sum_{n'} \arrowvert P_{n,n'}(k_m) \arrowvert^2
\\ && \ \ \ \ \ \times
\left( f_{n'}(k_m) - f_n(k_m) \right)
\Delta_{n,n'}^m(\hbar\omega)
\label{Photogeneration term}
\end{eqnarray}
The cell averaged delta function, denoted $\Delta_{n,n'}^m(\varepsilon)$, is
\begin{equation}
\Delta_{n,n'}^m(\varepsilon) =
\int_m \frac{dk \ k}{\Delta k \ k_m} \ \delta_\gamma \left( \arrowvert
E_{n'}(k) - E_{n}(k) \arrowvert - \varepsilon \right)
\end{equation}
where $\delta_\gamma(x)$ is a Lorentzian lineshape with FWHW, $\gamma$.

The squared optical matrix element $\arrowvert P_{n,n'}(k) \arrowvert^2$
between subbands $n$ and $n'$ is the angular average
\begin{equation}
\arrowvert P_{n,n'}(k) \arrowvert^2
= \int_{-\pi}^{\pi} \frac{d \theta}{2 \pi} \ \arrowvert \
\mathbf{\epsilon} \cdot \mathbf{P}_{n,n'}(k,\theta) \ \arrowvert^2
\end{equation}
where $\mathbf{\epsilon}$ is the unit complex polarization vector of the
pump pulse. The optical matrix element between subband states at fixed
$\mathbf{k}_\parallel$ is
\begin{equation}
\mathbf{P}_{n,n'}(\mathbf{k}_\parallel)=\frac{\hbar}{m_0} \sum_{\nu,\nu'}
\langle \nu \arrowvert \mathbf{p} \arrowvert \nu' \rangle
\int dz \ F_{n,\nu,\mathbf{k}_\parallel}^{*}(z)
F_{n',\nu',\mathbf{k}_\parallel}(z)
\label{Optical matrix elements}
\end{equation}
where $\langle \nu \arrowvert \mathbf{p} \arrowvert \nu' \rangle$ is
the momentum matrix element between the Bloch basis states defined
in Eqs. (\ref{upperset}) and (\ref{lowerset}). Explicit expressions
for the matrix elements of $\mathbf{P} = (\hbar/m_0) \ \mathbf{p}$
between the Bloch basis states in terms of the Kane momentum matrix
element, $V$, defined in Eq. (\ref{Kane momentum matrix element})
can be found in Appendix B of Ref. \onlinecite{Sanders03.165205}.

\subsection{Recombination rates due to carrier trapping}

Electron-hole pairs can recombine through the ultrafast trapping of
electrons (by As$_{Ga}$ antisite defects) and holes (by Ga
vacancies) by the mid-gap states introduced by LT-MBE growth
\cite{Wang05.0505261, Wang05.accepted}. The cell averaged carrier
trapping rate in Eq. (\ref{Photogeneration and Relaxation Terms}) is
treated using a simple relaxation time model of the form
\begin{equation}
\left[ \frac{\partial f_n(k_m)}{\partial t} \right]_T =
- \left( \frac{f_n(k_m,t)-f^0_n(k_m)}{ \tau(t) } \right)
\label{Carrier trapping term}
\end{equation}
where $f^0_n(k_m)$ are the initial thermal equilibrium distribution
functions defined in Eq. (\ref{Initial distribution function}) and $\tau(t)$
is the Shockley-Read recombination time for electron-hole pairs.

If we assume electron-hole pairs recombine through trapping at monovalent
flaws in the mid-gap region, the Shockley-Read recombination time can be
expressed as \cite{Blakemore}
\begin{equation}
\tau(t) = \frac{(n+p) \ \tau_0 +n_e(t) \ \tau_\infty}{(n+p)+n_e(t)}
\label{Shockely Read lifetime}
\end{equation}
where $n$ and $p$ are the initial electron and hole column densities in the
quantum well and $n_e(t)$ is the column density of photogenerated electron-hole
pairs. In our model, we assume for simplicity that the flaws are
acceptor-like with $\tau_\infty \ll \tau_0$ so that $\tau_\infty \approx 0$.

The electron-hole pair column density is equal to the column density of
\emph{photogenerated} electrons which is given by
\begin{equation}
n_e(t) = \frac{1}{2\pi} {\sum_{n,m}}' \Delta k \ k_m
\left( \  f_n(k_m,t) - f^0_n(k_m) \ \right)
\label{Electron-hole pair density}
\end{equation}
where the prime on the summation sign is a reminder that the sum over
subband index, $n$, is restricted to conduction subbands. We note in
passing that the number of photogenerated electrons and holes remain
equal in the Shockley-Read recombination model.

\subsection{Confined LO phonon scattering rates}

When the sample is excited by the ultrafast pump laser, hot carriers
are created above the fundamental gap. The hot carriers relax
back to the band edge and reach a quasi-thermal equilibrium through
carrier cooling. The dominant cooling mechanism is absorption
and emission of confined longitudinal optical (LO) phonons in the quantum
well. The Fr\"{o}lich Hamiltonian for LO phonon scattering in a quantum well
of width $L$ is given by \cite{Sanders98.9148}
\begin{equation}
H_{LO}=\frac{C_{LO}}{\sqrt{L A}} \sum_{\mathbf{q},l}
t_l(q) \ e^{i \mathbf{q} \cdot \mathbf{\rho}} \ u_l(z)
\left( a_{\mathbf{q},l}^\dag + a_{-\mathbf{q},l} \right),
\end{equation}
where the electron-LO phonon coupling constant is given by
\begin{equation}
C_{LO}= \sqrt{ 4 \pi \ \hbar \omega_{LO} \left(
\frac{1}{\varepsilon_\infty} - \frac{1}{\varepsilon_0} \right) }.
\end{equation}
The LO phonon energy in the Einstein model is $\hbar \omega_{LO}$ and
$\varepsilon_0$ and $\varepsilon_\infty$ are the static
and high frequency dielectric constants in the quantum well.
The operator $a_{\mathbf{q},l}^\dag$ creates a confined LO phonon
in the quantum well in the $l$th LO phonon mode with wavevector $\mathbf{q}$.
The vibrational amplitude of the $l$th LO phonon mode is $u_l(z)$
and
\begin{equation}
\frac{1}{t_l(q)^2}=\frac{2}{L} \int_{0}^{L} dz
\left[ q^2 u_l(z)^2+ \left(
\frac{\partial u_l(z)}{\partial z}
\right)^2 \right].
\end{equation}
The vibrational amplitude is model dependent. In the slab mode
model of Fuchs and Kleiwer \cite{Fuchs65.2076}
\begin{equation}
u_l(z)= \sin \left( \frac{l \pi z}{L} \right) \ \ \ l=1,2,3 \ \ldots
\end{equation}

Using Fermi's golden rule, the confined LO phonon scattering rate
due to emission or absorption of a single confined LO phonon is
\begin{widetext}
\begin{equation}
W_{\mathbf{k}_\parallel,\mathbf{k}_\parallel'}^{n,n'}=
\frac{2\pi}{\hbar} \frac{C_{LO}^2}{LA} \sum_{l=1}^{\infty}
t_l ( \arrowvert\mathbf{k}_\parallel-\mathbf{k}_\parallel' \arrowvert)^2
\ \arrowvert
\gamma_{\mathbf{k}_\parallel,\mathbf{k}_\parallel'}^{n,n'}(l)
\arrowvert^2 \left[
N_0 \ \delta \left( \Delta E^{n,n'}_{k,k'}+\hbar\omega_{LO} \right)
+(N_0+1)\ \delta \left( \Delta E^{n,n'}_{k,k'}-\hbar\omega_{LO} \right)
\right].
\label{LO scattering rate}
\end{equation}
\end{widetext}
The energy slitting $\Delta E^{n,n'}_{k,k'}=E_n(k)-E_{n'}(k')$ and the
LO phonon occupation number, $N_0$, is given by the Bose-Einstein
distribution
\begin{equation}
N_0 = \frac{1}{\exp (\hbar\omega_{LO}/k_B T)-1}.
\end{equation}
The vibrational amplitude form factor is defined as
\begin{equation}
\gamma_{\mathbf{k}_\parallel,\mathbf{k}_\parallel'}^{n,n'}(l)=
\sum_{\nu=1}^8 \int dz \ F_{n,\nu,\mathbf{k}_\parallel}^{*}(z)
\ u_l(z) \ F_{n',\nu,\mathbf{k}_\parallel'}(z)
\end{equation}
The cell averaged scattering rates, $W_{m,m'}^{n,n'}$
used in the Bolztmann equation (\ref{1D Boltzmann equation}) are
obtained by substituting Eq. (\ref{LO scattering rate})
into Eq. (\ref{Cell averaged scattering rates}) and performing
the integrals.

\subsection{Generation and propagation of coherent acoustic phonons}

The ultrafast photogeneration of electrons and holes in the InMnAs
quantum well by the pump gives rise to coherent longitudinal
acoustic (LA) phonons which propagate into the sample.
\cite{Sanders01.235316,Sanders02.079903,Chern04.339,Stanton03.525,Yahng02.4723}
Coherent acoustic phonons, as opposed to incoherent phonons, give rise
to a macroscopic lattice displacement. Since the photogenerated
carrier distributions are functions of $z$, the
transient lattice displacement $U(z,t)$ due to photogenerated carriers
is independent of $x$ and $y$ and is parallel to $z$. As discussed in
Refs. \onlinecite{Sanders01.235316} and \onlinecite{Chern04.339},
the coherent phonon lattice displacement satisfies a loaded string
equation. In the presence of a position dependent longitudinal acoustic
sound velocity, $C_s(z)$, we have
\begin{equation}
\frac{\partial^2 U(z,t)}{\partial t^2} -
\frac{\partial}{\partial z}
\left( C_s(z)^2 \ \frac{\partial U(z,t)}{\partial z}
\right)
=S(z,t)
\label{Loaded string equation}
\end{equation}
where $S(z,t)$ is the driving function. The longitudinal
acoustic sound velocity is given by
\begin{equation}
C_s(z) = \sqrt{\frac{C_{11}(z)}{\rho_0(z)}}
\label{Sound velocity}
\end{equation}
where $C_{11}(z)$ and $\rho_0(z)$ are the position dependent
elastic stiffness constant and mass density.

The loaded string
equation is to be solved subject to the initial conditions
\begin{equation}
U(z,t=-\infty) = \frac{\partial U(z,t=-\infty)}{\partial t} = 0.
\label{U(z,t) initial conditions}
\end{equation}
We solve the loaded string equation numerically by finite
differencing Eq. (\ref{Loaded string equation}) inside a
computational box whose left edge, $z_L$, is the semiconductor-air
interface and whose right edge, $z_R$, lies inside the GaAs substrate
(see Fig. \ref{Structure}).
At $z_R$ we impose absorbing boundary conditions and at $z_L$ there
are no perpendicular forces at the semiconductor-air interface.
Thus we solve the initial value problem subject to the left
and right boundary conditions
\begin{subequations}
\label{U(z,t) boundary conditions}
\begin{equation}
\frac{\partial U(z_L,t)}{\partial z} = 0
\end{equation}
and
\begin{equation}
\frac{\partial U(z_R,t)}{\partial z}+
\frac{1}{C_s(z_R)} \ \frac{\partial U(z_R,t)}{\partial t}=0.
\end{equation}
\end{subequations}

Starting with the second quantized Hamiltonian for the electron-phonon
interaction, a microscopic expression for the driving function was
derived in Ref. \onlinecite{Sanders01.235316} using the density
matrix formalism (see also the erratum in
Ref. \onlinecite{Sanders02.079903} as well as the review
article in Ref. \onlinecite{Chern04.339}).
In zinc-blende materials such as InAs the electron-phonon
interaction is due to deformation potential coupling. Under typical
experimental conditions the microscopic expression for the driving
function can be simplified to \cite{Sanders01.235316}
\begin{equation}
S(z,t) = \frac{1}{\rho_0}
\left(
a_c \frac{\partial \rho_e(z,t)}{\partial z}-
a_v \frac{\partial \rho_h(z,t)}{\partial z}
\right)
\label{Simplified S(z,t)}
\end{equation}
where $\rho_0$ is the mass density, $a_c$ and $a_v$ are the
deformation potentials for conduction and valence bands and
$\rho_e(z,t)$ and $\rho_h(z,t)$ are the photogenerated electron
and hole carrier densities. We note that this last equation was
derived independently in the elastic continuum limit by
Chigarev et al \cite{Chigarev00.15837} and
Chern et al. \cite{Chern04.339} The driving function satisfies
the sum rule
\begin{equation}
\int_{-\infty}^{\infty} dz \ S(z,t) = 0.
\label{Sum rule}
\end{equation}
as shown in Refs. \onlinecite{Sanders01.235316} and
\onlinecite{Chern04.339}.

The photogenerated electron and hole
densities can be obtained from the envelope and
distribution functions described in the previous sections
as
\begin{subequations}
\label{Photogenerated carrier densities}
\begin{equation}
\rho_e(z,t)=\frac{1}{A} \sum_{n,\nu,\mathbf{k}_\parallel}
\left( f_n(k,t)-f_n^0(k) \right)
\left| F_{n,\nu,\mathbf{k}_\parallel}(z) \right|^2 \delta_{n,c}
\end{equation}
and
\begin{equation}
\rho_h(z,t)=\frac{1}{A} \sum_{n,\nu,\mathbf{k}_\parallel}
\left( f_n^0(k)-f_n(k,t) \right)
\left| F_{n,\nu,\mathbf{k}_\parallel}(z) \right|^2 \delta_{n,v}.
\end{equation}
\end{subequations}
The initial Fermi-Dirac distribution functions $f_n^0(k)$ are
defined in Eq. (\ref{Initial distribution function}) and
$\delta_{n,c}$ and $\delta_{n,v}$ are
conduction and valence delta functions that select
out conduction and valence subbands, respectively.

The propagating coherent phonon displacement field $U(z,t)$ gives
rise to a propagating strain tensor with components
\begin{subequations}
\label{Strain tensor components}
\begin{equation}
\varepsilon_{zz}(z,t)=\frac{\partial U(z,t)}{\partial z}
\end{equation}
and
\begin{equation}
\varepsilon_{xx}(z,t)=\varepsilon_{yy}(z,t)=
- \ \frac{C_{12}(z)}{C_{11}(z)+C_{12}(z)} \ \varepsilon_{zz}(z,t)
\end{equation}
\end{subequations}
where $C_{11}(z)$ and $C_{12}(z)$ are elastic stiffness constants.
This propagating strain field alters the optical properties of the
sample which can be detected by the probe.

\subsection{Transient probe response}

\begin{figure}[tbp]
\includegraphics[scale=0.7]{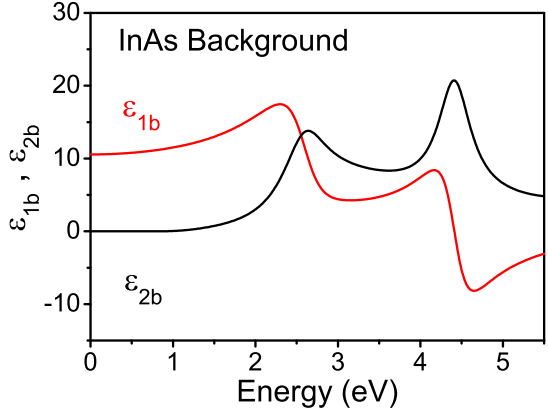}
\caption{Background model dielectric function at $T = 0 \ \mbox{K}$
as a function of photon energy
used in computing the dielectric function in the InMnAs quantum well.}
\label{InAs background dielectric function}
\end{figure}

To compute the time dependent probe transmission and reflection
coefficients we need to model the dielectric function in the entire
structure, i.e. the InMnAs well, GaSb barrier, and  GaAs substrate,
over the probe energy range which extends up to 2 eV. We will
denote this dielectric function as
\begin{equation}
\varepsilon(\hbar\omega,z,t) =
\varepsilon_1(\hbar\omega,z,t) + i \ \varepsilon_2(\hbar\omega,z,t)
\label{Total dielectric function}
\end{equation}
where $\hbar \omega$ is the probe energy. Once the dielectric function
is found, we can solve Maxwell's equations for the time dependent probe
reflection coefficient, $R(\hbar\omega,t)$, and the transmission
coefficient, $T(\hbar\omega,t)$, using the transfer matrix method
described in detail in Ref. \onlinecite{Chuang}.

There are several processes which contribute to the dielectric function
in the InMnAs quantum well. The first of these is a Drude term
due to free carriers in the quantum well which gives a real contribution
to the dielectric function of
\begin{equation}
\varepsilon_1(\hbar\omega,z,t)_{D}=
-\frac{\left(\hbar\omega_p(t) \right)^2}{\left( \hbar\omega \right)^2}
\label{Drude term}
\end{equation}
where $\omega_p$ is the plasma frequency. The Drude contribution
to the dielectric function in Eq. (\ref{Drude term}) is uniform
in the quantum well and vanishes everywhere else. In the random phase
approximation (RPA), the time dependent plasma frequency is given
by \cite{Schafer}
\begin{equation}
\omega_p^2(t)=\frac{4 \pi e^2}{L}
\frac{1}{A} \sum_{n,\mathbf{k}_\parallel}
\frac{f_n(k,t)-\delta_{n,v}}{m_n^*(k)}
\label{Plasma frequency}
\end{equation}
where $L$ is the quantum well width and $A$ is the
cross sectional area of the sample. The effective mass is
\begin{equation}
\frac{1}{m_n^*(k)}=\frac{1}{\hbar^2} \
\frac{\partial^2 E_n(k)}{\partial k^2}.
\label{Plasma frequency effective mass}
\end{equation}

A second contribution to the dielectric function in the quantum
well is due to dipole transitions between quantum confined carrier
states. Using Fermi's golden rule, we obtain \cite{Bassani,Chuang}
\begin{widetext}
\begin{subequations}
\label{Quantum well dielectric function}
\begin{equation}
\varepsilon_1(\hbar\omega,z,t)_{QW} =
\frac{8 \pi e^2}{L}
\frac{1}{A} \sum_{n,n',k_\parallel}
\arrowvert P_{n,n'}(k_\parallel) \arrowvert^2
\frac{f_{n'}(k,t)-f_n(k,t)}
{\Delta E_{n,n'}(k) \left( \Delta E_{n,n'}(k)+\hbar\omega \right)
\left( \Delta E_{n,n'}(k)-\hbar\omega\right)}
\end{equation}
and
\begin{equation}
\varepsilon_2(\hbar\omega,z,t)_{QW} =
\frac{4 \pi^2 e^2}{L \ (\hbar\omega)^2}
\frac{1}{A} \sum_{n,n',k_\parallel}
\arrowvert P_{n,n'}(k_\parallel) \arrowvert^2
\left( f_{n'}(k,t)-f_n(k,t) \right)
\ \delta( \Delta E_{n,n'}(k)-\hbar\omega )
\end{equation}
\end{subequations}
\end{widetext}
where $\Delta E_{n,n'}(k) = E_n(k) - E_{n'}(k)$ are the
transition energies, including band gap renormalization corrections
due to photogenerated carriers, and  $P_{n,n'}(k_\parallel)$ are the
optical matrix elements. The contributions to the dielectric function
in Eq. (\ref{Quantum well dielectric function}) are for zero linewidth.
For a finite FWHM linewidth $\Gamma$, we make the replacements
\cite{Chuang}
\begin{equation}
\frac{1}{\Delta E_{n,n'}(k)-\hbar\omega} \rightarrow
\frac{\Delta E_{n,n'}(k)-\hbar\omega}
{ \left( \Delta E_{n,n'}(k)-\hbar\omega \right)^2 +(\Gamma/2)^2 }
\end{equation}
and
\begin{equation}
\delta(x) \rightarrow
\frac{1}{\pi} \ \frac{(\Gamma /2)}{x^2+(\Gamma / 2)^2}
\end{equation}

There is also a background dielectric function, $\varepsilon_b$ in the
quantum well due to all the higher lying electronic transitions whose
real and imaginary parts we shall denote $\varepsilon_{1b}$ and
$\varepsilon_{2b}$. For simplicity, we treat these contributions to the
dielectric function using Adachi's model dielectric function for bulk
InAs with contributions from the $E_0$ and $E_0+\Delta_0$ critical points
removed. \cite{Adachi89.6030,Djurisic99.3638} These correspond to
contributions from the confined quantum well electronic states and are
already included in Eq. (\ref{Quantum well dielectric function}). The
background dielectric function in InAs as a function of photon energy is
shown in Figure \ref{InAs background dielectric function} at
$T = 0 \ \mbox{K}$ and in the absence of strain.

Following Thomsen et. al.\cite{Thomsen84.989,Thomsen86.4129}
we assume that the dielectric function changes with strain only because
of strain induced variations in the energy gaps associated with  each transition.
The propagating coherent phonon strain tensor alters the optical properties
of the structure through the deformation potential interaction.
In our experiments, the probe photon energy can go as high as the
GaSb $E_1$ transition region.
Ab initio density functional calculations of the deformation potentials
for the $E_1$ transitions in a number of semiconductors \cite{Ronnow99.5575}
have shown that the deformation potentials associated with the $E_0$ and $E_1$
features are equal to within 20\%. So to a first approximation the effect of
temperature and strain on $\varepsilon_b$ is to introduce a rigid shift in the
dielectric function such that
\begin{equation}
\varepsilon_{b}(\hbar\omega,z,t)=
\varepsilon_{b}(\hbar\omega-\Delta E_g(T)-
a_{cv}(\varepsilon_{xx}+\varepsilon_{yy}+\varepsilon_{zz})).
\label{Background dielectric function shift}
\end{equation}
Here, $a_{cv}=a_c-a_v$ is the interband deformation potential,
$\varepsilon_{xx}$, $\varepsilon_{yy}$, and $\varepsilon_{zz}$ are the
coherent phonon strain tensor components defined in
Eq.~(\ref{Strain tensor components}),
and $\Delta E_g(T) = E_g(T)-E_g$ is the band gap shift due to
temperature variations with $E_g(T)$ being the temperature dependent
band gap defined by the Varshni expression in Eq. (\ref{Varshni formula}).

The total dielectric function in the quantum well is obtained by adding
the Drude, Quantum well, and background contributions in
Eqs.(\ref{Drude term}), (\ref{Quantum well dielectric function}) and
(\ref{Background dielectric function shift}), i.e.
\begin{equation}
\varepsilon(\hbar\omega,z,t)=
\varepsilon(\hbar\omega,z,t)_{D}
+\varepsilon(\hbar\omega,z,t)_{QW}
+\varepsilon_{b}(\hbar\omega,z,t)
\label{Sum total of dielectric functions}
\end{equation}
%

\begin{figure}[tbp]
\includegraphics[scale=0.9]{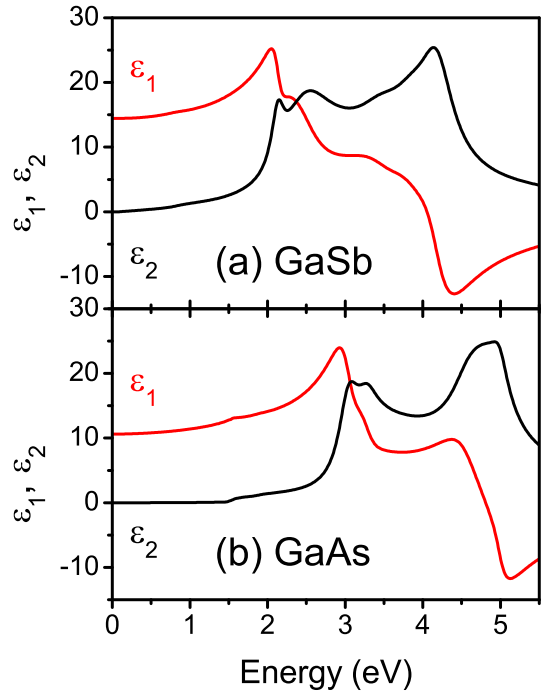}
\caption{Model dielectric functions at $T = 0 \ \mbox{K}$
as a function of photon energy for (a) GaSb
and (b) GaAs used in calculating the dielectric functions in the GaSb
barrier and GaAs substrate.}
\label{Model dielectric functions}
\end{figure}

We use Adachi-type model dielectric functions for the GaSb barrier
\cite{Munoz99.8105,Djurisic01.902}
and the GaAs substrate. \cite{Rakic96.5909}
Fig. \ref{Model dielectric functions} shows the real and imaginary
parts of the model dielectric function for bulk GaSb and GaAs at
$T = 0 \ \mbox{K}$ in the absence of strain. Temperature and
strain effects in the GaSb barrier and GaAs substrate are included
using the same rigid shift model
as defined in Eq. (\ref{Background dielectric function shift}).
Note that the dielectric functions in GaSb and GaAs are modulated
by the coherent phonon strain field as it propagates through the
structure.

\section{Results and discussion}

\subsection{Quantum well electronic states}

\begin{figure}[tbp]
\includegraphics[scale=0.82]{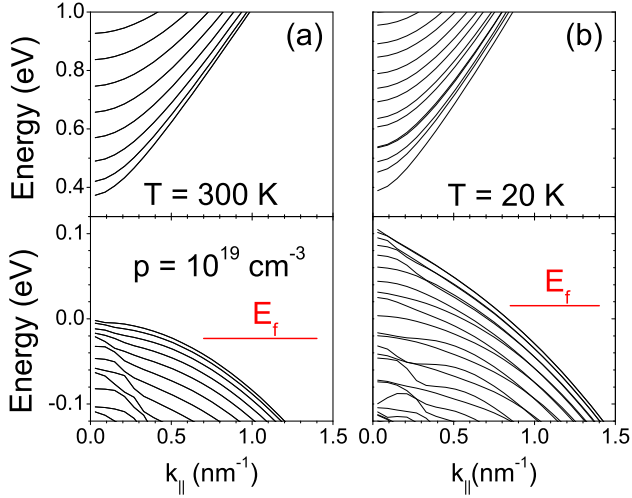}
\caption{ Bandstructure of a 25 nm ferromagnetic In$_{1-x}$Mn$_x$As
quantum well with infinite barriers and $x = 9 \ \%$ at
(a) $T = 300 \ \mathrm{K}$ and (b) $T = 20 \ \mathrm{K}$.
The Curie temperature is taken to be $T_c = 55 \ \mathrm{K}$.
The Fermi energies, $E_f$, for a hole density of
$p=10^{19} \mathrm{cm}^{-3}$ are indicated by the horizontal lines.}
\label{Bandstructure}
\end{figure}

The 25 nm thick ferromagnetic In$_{0.91}$Mn$_{0.09}$As quantum well is
shown schematically in Fig.~\ref{Structure}. The quantum well is p-type
with a free hole density estimated to be
$p \approx 10^{19} \ \mathrm{cm}^{-3}$ and a Curie temperature of
$T_c = 55 \ \mathrm{K}$. The computed bandstructure of the 25 nm
ferromagnetic In$_{0.91}$Mn$_{0.09}$As quantum well in the axial
approximation, assuming infinite barriers, is shown in
Fig.~\ref{Bandstructure} as a function of $k_\parallel$ for
temperatures above and below the Curie temperature.
The Fermi energies, $E_f$, corresponding to a free hole density of
$p = 10^{19} \ \mathrm{cm}^{-3}$ are indicated by short horizontal
lines in Fig.~\ref{Bandstructure}.
The bandstructure well above the Curie temperature
($T = 300 \ \mathrm{K}$) is shown in Fig.~\ref{Bandstructure}(a).
Far above the Curie temperature, the average Mn spin
$\langle S_z \rangle$ vanishes, the sample is nonmagnetic, and the
computed subbands are doubly degenerate. Below the Curie temperature,
the quantum well becomes ferromagnetic with a nonvanishing
$\langle S_z \rangle$ and the doubly degenerate subbands
in Fig.~\ref{Bandstructure}(a) become
spin-split as can be seen in Fig.~\ref{Bandstructure}(b).

\subsection{Coherent phonon generation and propagation}

\begin{figure}[tbp]
\includegraphics[scale=0.82]{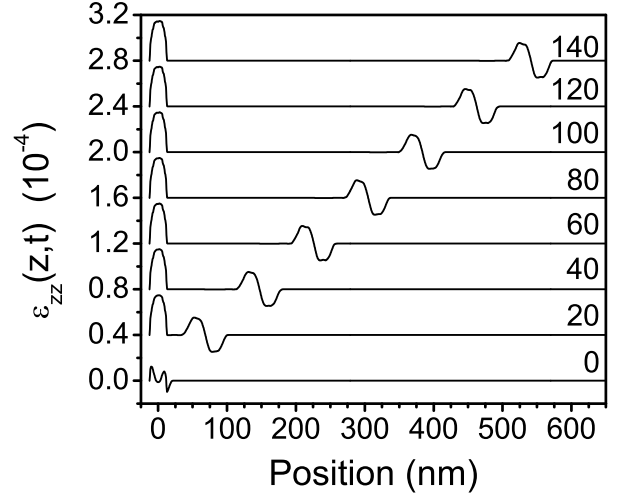}
\caption{ Coherent phonon strain field as a function of position
for delay times ranging from 0 to 140 ps. The curves have been offset
for clarity. }
\label{Propagating strain}
\end{figure}

The coherent phonon lattice displacement, $U(z,t)$, obtained from
the loaded string equation (\ref{Loaded string equation}) gives rise to
strain tensor components $\varepsilon_{xx}(z,t)$, $\varepsilon_{yy}(z,t)$,
and $\varepsilon_{zz}(z,t)$ as defined in
Eq.~(\ref{Strain tensor components}). The  coherent phonon strain tensor
component $\varepsilon_{zz}(z,t)$ is shown in
Fig.~\ref{Propagating strain} as a function of position in the sample
for several equally spaced delay times ranging from 0 to 140 ps.

Following photogeneration of carriers by the pump, a localized strain
appears in the quantum well as can be seen in Fig.~\ref{Propagating strain}.
This is due to near steady-state loading by the driving function at long
times. Assuming the driving function, $S(z,t)$, is approximately  time
independent at long times, the loaded string equation
(\ref{Loaded string equation}) can be integrated once in the steady-state
limit. The resulting steady-state strain is
\begin{equation}
\varepsilon_{zz}(z)=\frac{\partial U(z)}{\partial z}=
-\int_{-\infty}^{z} dz' \ \frac{S(z')}{C_s^2}.
\label{Steady state strain}
\end{equation}
where $C_s$ is the longitudinal acoustic sound speed in the InMnAs
quantum well and $S(z')$ is the approximately time independent driving
function left behind in the quantum well at long times.
The fact that the steady-state strain is localized in the
well follows directly from the sum rule (\ref{Sum rule}).

In addition to the localized strain, a transient strain pulse propagates
into the GaSb barrier at the longitudinal acoustic sound speed. Two
transient strain pulses are generated in the well, one propagating to the
left and the other to the right. The leftward propagating pulse is totally
reflected off the semiconductor-air interface and trails the rightward
propagating pulse as it propagates into the GaSb barrier.

\subsection{Differential reflectivity}

The oscillation observed in the differential reflectivity can be
attributed to propagation of the strain pulse through the structure.
During most of the experiment, the travelling strain pulse is in
the GaSb barrier shown schematically in Fig. \ref{Structure}.

The propagating strain tensor shown in Fig.~\ref{Propagating strain}
alters the dielectric function in the structure. The change in the
complex dielectric function due to coherent phonon wavepackets
is given by
\begin{equation}
\Delta \varepsilon(\hbar\omega,z,t)=
\frac{d \varepsilon(\hbar\omega,z)}{d \varepsilon_{zz}}
\ \varepsilon_{zz}(z,t),
\label{Dielectric function derivatives wrt strain}
\end{equation}
where the total derivative with respect to strain,
$d \varepsilon(\hbar\omega,z)/d \varepsilon_{zz}$,
is piecewise constant in $z$ having different values in
the InMnAs well, the GaSb barrier, and the GaAs substrate.

\begin{figure}[tbp]
\includegraphics[scale=0.75]{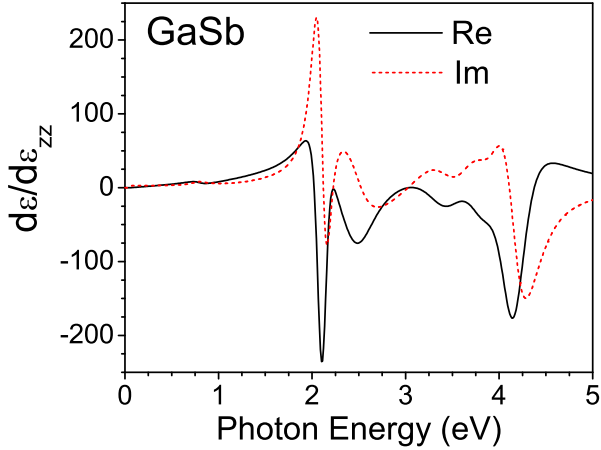}
\caption{ Derivative of the complex GaSb dielectric function with
respect to strain as a function of the probe photon energy. The solid
line is the real part and the dashed line is the imaginary part.}
\label{GaSb dielectric function derivatives wrt strain}
\end{figure}

The total derivative with respect to strain in
Eq.~(\ref{Dielectric function derivatives wrt strain}) is
obtained from the Adachi-type model dielectric functions by
differentiating Eq. (\ref{Background dielectric function shift}) with
respect to $\varepsilon_{zz}$ taking care to eliminate
$\varepsilon_{xx}$ and $\varepsilon_{yy}$ in favor of $\varepsilon_{zz}$
using Eq.~(\ref{Strain tensor components}). In the case of GaSb, the
real and imaginary parts of $d \varepsilon /d \varepsilon_{zz}$
are plotted as a function of the probe photon energy in
Fig.~\ref{GaSb dielectric function derivatives wrt strain}.
%
%
As seen in Fig.~\ref{GaSb dielectric function derivatives wrt strain}
the perturbation in the dielectric function per unit strain, i.e.
the dielectric function strain sensitivity, is a function of the probe
energy. In particular, there is a giant peak in the dielectric function
strain sensitivity at the GaSb $E_1$ transition near 2.0 eV.
Consequently, the best probe wavelength for observing coherent phonon
differential reflectivity oscillations is in the region around the
$E_1$ transition.

The differential reflectivity at a given probe delay time is obtained
from the dielectric function by solving Maxwell's equations in the
entire structure using the transfer matrix formalism as described
earlier. If we use the complete time- and space-dependent dielectric function
defined in Eq.~(\ref{Sum total of dielectric functions}) including the
quantum well, Drude, and background contributions, we can compute the
total differential reflectivity containing the coherent phonon
oscillation and transient carrier relaxation effects.
%
\begin{figure}[tbp]
\includegraphics[scale=.95]{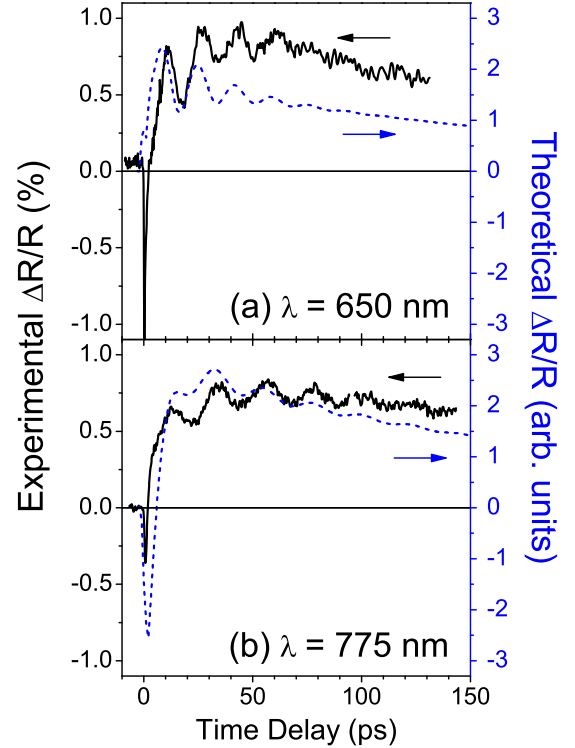}
\caption{Theoretical and experimental differential reflectivity for
probe wavelengths of (a) 650 and (b) 775 nm due to variations in the
time- and space-dependent dielectric functions. }
\label{Theoretical differential reflectivity}
\end{figure}
%

Our theoretical results are compared with experiment in
Fig.~\ref{Theoretical differential reflectivity} where we plot
the experimental and theoretical differential reflectivity spectra for
probe wavelengths of 650 and 775 nm. In both cases there is an initial sharp
drop in the differential reflectivity which we attribute to free carrier
Drude absorption by the hot carriers created by the pump.

The photogenerated hot carriers relax back to quasi-equilibrium distributions
at their respective band edges through emission of confined LO phonons.
The relaxation of photogenerated carriers by LO phonons alters the
quantum well dielectric function in
Eq.~(\ref{Quantum well dielectric function}) through changes in the
time-dependent distribution functions. This carrier cooling by LO phonon
emission results in the subsequent rise in the differential reflectivity
traces seen if Fig.~\ref{Theoretical differential reflectivity}.

In addition to carrier cooling by LO phonon emission, electron-hole pairs
recombine through trapping at mid-gap defects with $\tau_0 \approx \mbox{200 ps}$.
This gives rise to the slow decay in the differential reflectivity at long
times seen in Fig.~\ref{Theoretical differential reflectivity}. At short
times, electron-hole pair recombination is enhanced since the Shockley-Read
recombination time, $\tau(t)$, is a monotonically decreasing function of
the photogenerated electron-hole pair density.

For delay times of less than 20 ps, our theory doesn't agree
very well with the experiment.
For delay times greater than 20 ps, however, the theory reproduces the
experimental results surprisingly well. In particular, the period and
amplitude of the reflectivity oscillations in relation to the height
of the plateau as well as the decay of the reflectivity oscillations
with delay time are in good agreement with experiment.

%
\begin{figure}[tbp]
\includegraphics[scale=.6]{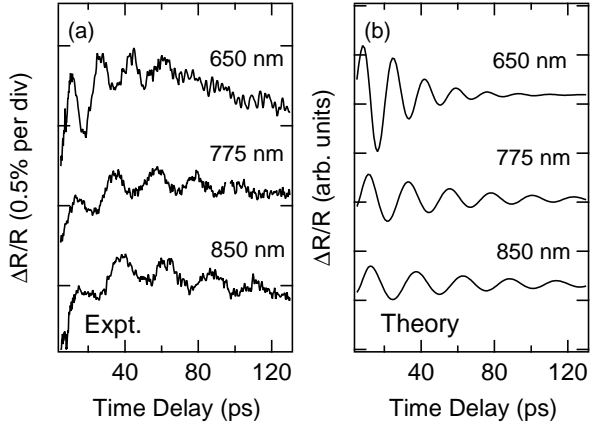}
\caption{Experimental (a) and Theoretical (b) coherent phonon
differential reflectivity oscillations for probe wavelengths of
650, 775 and 850 nm.}
\label{Coherent Phonons Theory vs Experiment}
\end{figure}

The oscillations in differential reflectivity seen in
Fig.~\ref{Theoretical differential reflectivity} are due to
changes in the background dielectric function induced by the
propagating coherent strain pulse seen in Fig.~\ref{Propagating strain}.
If we compute the probe differential reflectivity neglecting the
quantum well and Drude contributions to the total dielectric function in
Eq.~(\ref{Sum total of dielectric functions}) and retain only the
background contribution, we get the coherent acoustic phonon
differential reflectivity oscillation absent the transient relaxation
signal.

In Fig.~\ref{Coherent Phonons Theory vs Experiment}
the computed coherent phonon differential reflectivity oscillations are
shown as a function of time delay for probe wavelengths of 650, 775 and
850 nm, corresponding to photon energies of 1.9, 1.6, and 1.46 eV
respectively. The theoretical differential reflectivity curves in
Fig.~\ref{Coherent Phonons Theory vs Experiment}~(b) agree well with
the experimentally measured differential reflectivity seen in
Fig.~\ref{Coherent Phonons Theory vs Experiment}~(a) after subtraction
of the transient background signal. As we go from 650 to 850 nm,
the differential reflectivity oscillation period becomes longer.

The reflectivity oscillations can be qualitatively understood as
follows. The propagating strain pulse in Fig.~\ref{Propagating strain}
gives  rise to a perturbation in the GaSb dielectric function which
propagates at the acoustic sound speed. The sample thus acts as a
Fabry-Perot interferometer and a simple geometrical optics argument
shows that the period for the reflectivity oscillations due to the
propagating coherent acoustic phonon wavepacket is
approximately \cite{Yahng02.4723}
\begin{equation}
T = \frac{\lambda}{2 \ C_s \ n(\lambda)}
\label{Oscillation period}
\end{equation}
where $\lambda=2\pi \ c/\omega$ is the probe wavelength, $C_s$ is the
LA sound speed in the GaSb barrier and $n(\lambda)$ is the wavelength
dependent refractive index. The refractive index can be obtained from
the GaSb model dielectric function in
Fig.~\ref{Model dielectric functions}~(a) as \cite{Haug}
\begin{equation}
n(\lambda)=\sqrt{ \frac{1}{2} \left(
\varepsilon_1(\lambda)+\sqrt{
\varepsilon_1(\lambda)^2 + \varepsilon_2(\lambda)^2}
\right) }
\label{Index of refraction}
\end{equation}
%
\begin{figure}[tbp]
\includegraphics[scale=0.85]{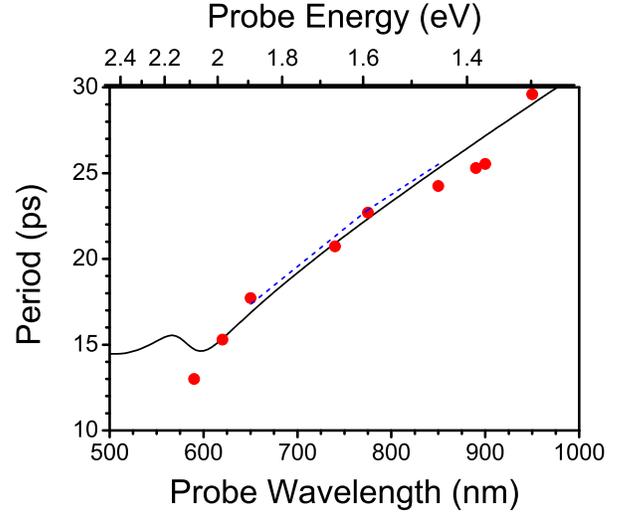}
\caption{ Coherent phonon differential reflectivity oscillation period
vs. probe wavelength. The experimental data are shown as solid circles,
the dashed line shows the oscillation period calculated from
the theory described in the text, and the solid line shows
the oscillation period estimated from Eq. \ref{Oscillation period}.}
\label{Experimental oscillation period vs wavelendth}
\end{figure}
%
In Fig.~\ref{Experimental oscillation period vs wavelendth} we have
plotted experimentally measured coherent phonon differential
reflectivity oscillation periods as a function of probe wavelength
as solid circles. The solid line shows the oscillation period
vs. probe wavelength estimated using Eq.~(\ref{Oscillation period}).
The excellent agreement between theory and experiment is compelling
evidence that the reflectivity oscillations seen in the experiments are
induced by propagating coherent acoustic phonons in the GaSb barrier.

\begin{figure}[tbp]
\includegraphics[scale=0.75]{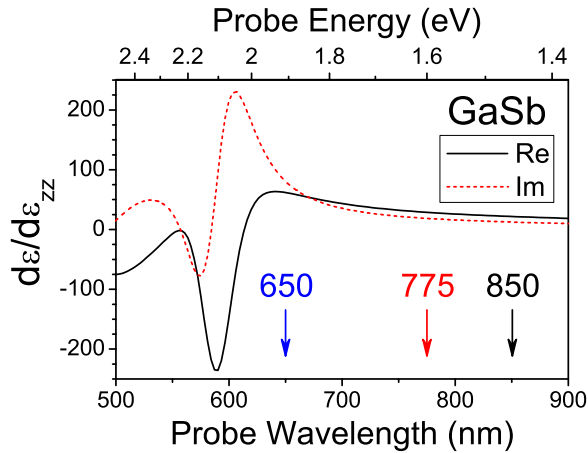}
\caption{ Derivative of the complex GaSb dielectric function with
respect to strain as a function of the probe wavelength in the
experimentally relevant wavelength range. The solid
line is the real part and the dashed line is the imaginary part.}
\label{GaSb wavelength dependent dielectric function derivatives wrt strain}
\end{figure}

In going from a probe wavelength of 650 to 850 nm in
Fig.~\ref{Coherent Phonons Theory vs Experiment},
we note that the initial amplitude of the
differential reflectivity oscillation decreases with increasing
probe wavelength. At the same time these
oscillations become more weakly damped. The reason for the reduction in
amplitude of the oscillations can be found in
Fig.~\ref{GaSb wavelength dependent dielectric function derivatives wrt strain}
where we plot $d \varepsilon /d \varepsilon_{zz}$ as a function of
probe wavelength. As the probe wavelength
increases (and the photon energy decreases), the strength of the
perturbation of the dielectric function due to the propagating
coherent phonon strain defined in
Eq.~(\ref{Dielectric function derivatives wrt strain}) decreases.  This
accounts for the observed reduction in the initial amplitude of the differential
reflectivity oscillations as we go to higher wavelengths.
The increased damping of the differential reflectivity oscillations with
decreasing probe wavelength is simply due to the fact that the
absorption coefficient in GaSb is rapidly decreasing with wavelength in
this wavelength range as can be inferred from the imaginary part of the
GaSb dielectric function plotted in
Fig.~\ref{Model dielectric functions}~(a).
%

\section{Conclusions}
\label{Conclusions}

In summary, we have performed calculations and modelled
time-dependent two-color differential reflectivity experiments on a
ferromagnetic InMnAs/GaSb heterostructure.  We have observed large
amplitude reflectivity oscillations resulting from the generation of
coherent acoustic phonon wavepackets in the InMnAs quantum well and
their subsequent propagation into the GaSb layer. The propagation of
these coherent, localized strain pulses into the GaSb buffer results
in a position- and frequency-dependent dielectric function.

To take into account the time dependent background differential
reflectivity, we modeled the two color pump-probe reflectivity
experiments in a Boltzmann equation formalism.  Electronic structure
was calculated using $\vec k \cdot \vec p$ theory in a confined
InMnAs layer. We included 1) photogeneration of hot carriers in the
InMnAs quantum well by a pump laser and 2) their subsequent cooling
by emission of confined LO phonons. Recombination of electron-hole
pairs via the Schockley-Read carrier trapping mechanism was also
included in a simple relaxation time approximation.

Our results agree remarkably well with the experimental coherent
phonon oscillations and reasonably well with the time-dependent
background signal and capture the major qualitative trends of the
data. We identify three key  effects which contribute to the
backgrounds signal: 1) the enhanced Drude absorption resulting from
the increase in carriers from the laser photoexcitation, (negative
$\Delta R/R$), 2)the relaxation dynamics associated with the decay
of the highly nonequilibrium photoexcited carrier distribution
(positive $\Delta R/R$) and 3) the trapping and then non-radiative
recombination of the photoexcited carriers resulting from the high
density of defects in the InMnAs layer (positive $\Delta R/R$).


\begin{acknowledgments}
This work was supported by  NSF through DMR-0134058 (CAREER),
DMR-0325474 (ITR),  and INT-0221704.
\end{acknowledgments}

\bibliography{paper}

\end{document}